\documentclass[twocolumn]{aastex631}

\hypersetup{linkcolor=magenta,citecolor=blue,filecolor=cyan,urlcolor=purple}

\received{2021 July 14}
\revised{2021 September 7}
\accepted{2021 September 10}

\shorttitle{Magnetic Field Prediction of a Stealth CME at PSP}
\shortauthors{Palmerio et al.}

\graphicspath{{./}{figures/}}

\turnoffeditone
\turnoffedittwo
\turnoffeditthree

\begin{document}

\title{Predicting the Magnetic Fields of a Stealth CME Detected by Parker Solar Probe at 0.5 AU}

\correspondingauthor{Erika Palmerio}
\email{epalmerio@berkeley.edu}

\author[0000-0001-6590-3479]{Erika Palmerio}
\affiliation{Space Sciences Laboratory, University of California--Berkeley, Berkeley, CA 94720, USA}
\affiliation{CPAESS, University Corporation for Atmospheric Research, Boulder, CO 80301, USA}

\author[0000-0002-2827-6012]{Christina Kay}
\affiliation{Heliophysics Science Division, NASA Goddard Space Flight Center, Greenbelt, MD 20771, USA}
\affiliation{Department of Physics, The Catholic University of America, Washington, DC 20064, USA}

\author[0000-0002-0973-2027]{Nada Al-Haddad}
\affiliation{Space Science Center, University of New Hampshire, Durham, NH 03824, USA}

\author[0000-0001-6886-855X]{Benjamin J. Lynch}
\affiliation{Space Sciences Laboratory, University of California--Berkeley, Berkeley, CA 94720, USA}

\author[0000-0002-2917-5993]{Wenyuan Yu}
\affiliation{Space Science Center, University of New Hampshire, Durham, NH 03824, USA}

\author[0000-0002-7728-0085]{Michael L. Stevens}
\affiliation{Smithsonian Astrophysical Observatory, Cambridge, MA 02138, USA}

\author[0000-0002-6302-438X]{Sanchita Pal}
\affiliation{Department of Physics, University of Helsinki, FI-00014 Helsinki, Finland}

\author[0000-0002-1604-3326]{Christina O. Lee}
\affiliation{Space Sciences Laboratory, University of California--Berkeley, Berkeley, CA 94720, USA}




\begin{abstract}

Stealth coronal mass ejection (CMEs) are eruptions from the Sun that are not associated with appreciable low-coronal signatures. Because they often cannot be linked to a well-defined source region on the Sun, analysis of their initial magnetic configuration and eruption dynamics is particularly problematic. In this \edit3{manuscript}, we address this issue by undertaking the first attempt at predicting the magnetic fields of a stealth CME that erupted in 2020~June from the Earth-facing Sun. We estimate its source region with the aid of off-limb observations from a secondary viewpoint and photospheric magnetic field extrapolations. We then employ the Open Solar Physics Rapid Ensemble Information (OSPREI) modelling suite to evaluate its early evolution and \edit2{forward-model}~its magnetic fields up to Parker Solar Probe, which detected the CME in situ at a heliocentric distance of 0.5~AU. We compare our hindcast prediction with in-situ measurements and \edit2{a set of}~flux rope reconstructions, obtaining encouraging agreement \edit2{on arrival time, spacecraft crossing location, and magnetic field profiles}. This work represents a first step towards reliable understanding and forecasting of the magnetic configuration of \edit2{stealth CMEs and slow, streamer-blowout}~events.

\end{abstract}



\section{Introduction} \label{sec:intro}

Eruptions of coronal mass ejections (CMEs) from the Sun are usually associated with a number of low-coronal signatures visible in solar disc imagery \citep{hudson2001}. These not only make it rather straightforward to identify where an eruption originates, but also allow for deeper analysis of a CME source region and estimation of the magnetic configuration of the corresponding flux rope \citep{palmerio2017}. Information about the internal magnetic structure of a CME in the solar corona at the time of its eruption is critical to being able to forecast CME magnetic fields in interplanetary space.

However, as was first reported by \citet{robbrecht2009}, there is a class of eruptions now known as ``stealth CMEs'' that lack the classic low-coronal signatures and hence are often more difficult to connect to a specific source region on the Sun. These events are usually slow and narrow, but they have been shown to occasionally drive significant geomagnetic disturbances at Earth \citep{nitta2017}. It follows that stealth CMEs, their magnetic fields, and their space weather effects are particularly problematic to forecast, especially because of the inability to observe their source and determine their flux rope configuration.

\begin{figure*}[!ht]
\includegraphics[width=\linewidth]{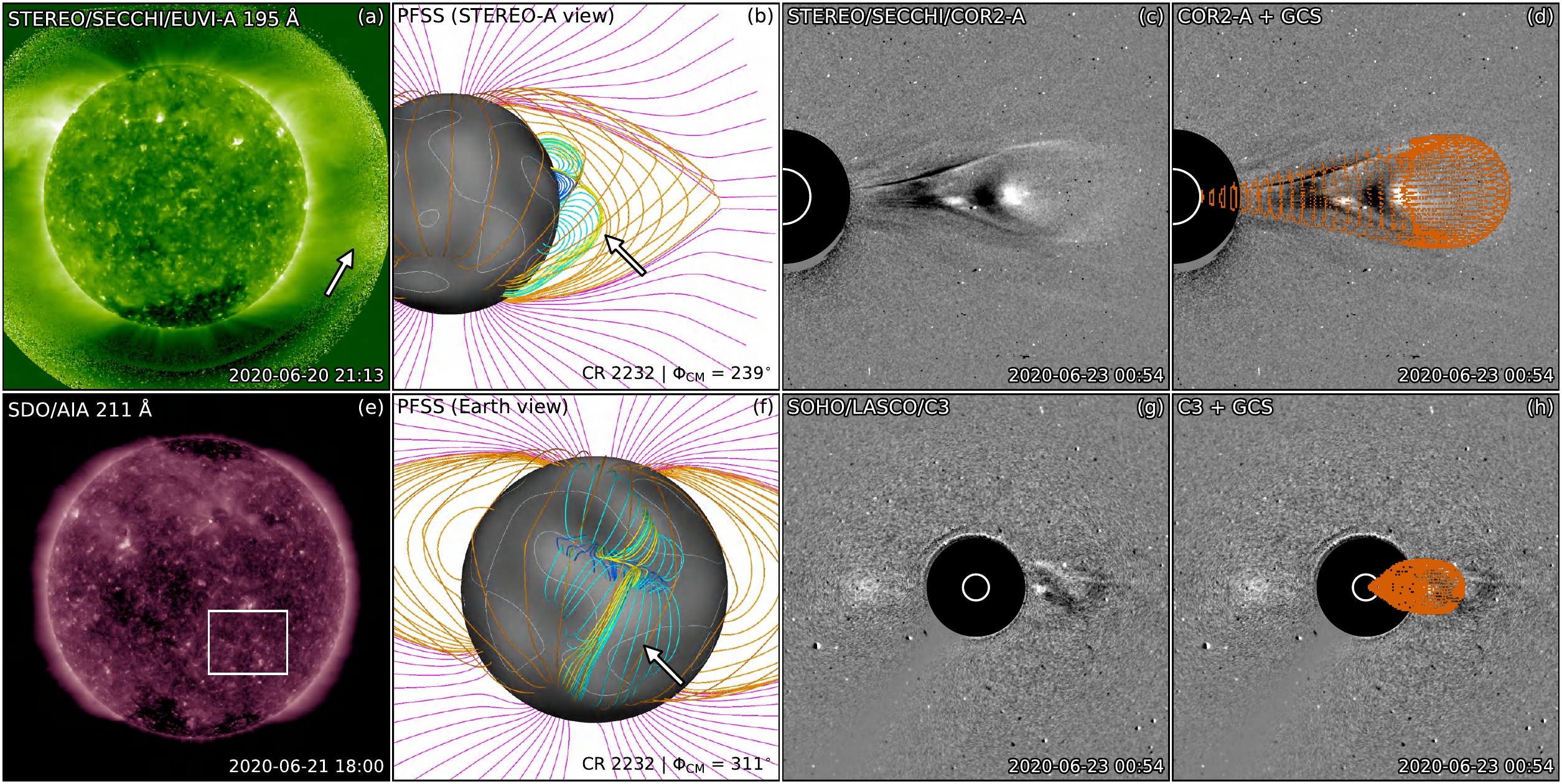}
\caption{Overview of the 2020~June~21 stealth CME eruption. (a) EUVI-A 195~{\AA} image showing the pre-eruptive flux rope structure (indicated with an arrow). (b) PFSS reconstruction \edit1{(with source at $2.5\,R_{\odot}$)} from the STEREO-A viewpoint showing the arcade involved in the eruption (indicated with an arrow). \edit1{The field lines are coloured by flux system connectivity (open polar fields: magenta; overlying helmet streamer: orange; side arcades: cyan; central arcade: blue; separatrices: yellow).} (c) COR2-A difference image showing the CME in white light and (d) with the GCS wireframe overlaid. (e) AIA 211~{\AA} image taken around the time of the CME eruption, showing an approximate source region (bounded by a box). (f) PFSS reconstruction from the Earth viewpoint \edit1{in the same colour scheme as (b)}, showing the PIL involved in the eruption (indicated with an arrow). (g) C3 difference image showing the CME in white light and (h) with the GCS wireframe overlaid. \label{fig:solar}}
\end{figure*}

In this work, we \edit2{present hindcast prediction results for} the magnetic fields of a stealth CME using remote-sensing observations and physics-driven \edit2{forward} models. The stealth CME that we focus on erupted on 2020~June~21 from the Earth-facing Sun and impacted the Parker Solar Probe \citep[PSP;][]{fox2016} spacecraft on June~25--26 at a heliocentric distance of 0.5~AU. We first provide an overview of the eruption and estimate the CME source region \edit2{location} based on off-limb observations and photospheric magnetic field extrapolations. We then use the Open Solar Physics Rapid Ensemble Information (OSPREI) suite of models to launch, propagate, and \edit2{predict the in-situ magnetic field profiles of} the stealth CME. Finally, we compare our prediction with PSP measurements, aided by flux rope reconstructions \edit2{characterising the} in-situ observations, and evaluate \edit2{the performance of the modelling chain used to self-consistently simulate the CME} from the Sun to 0.5~AU.

\section{Overview of the Eruption} \label{sec:solar}

The eruption that we analyse in this work initiated on 2020~June~21. An overview of remote-sensing observations from available imagery is shown in Figure~\ref{fig:solar}. The CME appeared as a narrow, slow streamer blowout (starting around 10:00~UT on June~22) in coronagraph data from the COR2 telescope (Figure~\ref{fig:solar}(c)) part of the Sun Earth Connection Coronal and Heliospheric Investigation \citep[SECCHI;][]{howard2008a} suite onboard the Solar Terrestrial Relations Observatory \citep[STEREO;][]{kaiser2008} Ahead spacecraft, located at ${\sim}1$~AU and ${\sim}70^{\circ}$ east of the Sun--Earth line at the time of the event. The CME was also observed by \edit3{the C2 and C3 telescopes} part of the Large Angle and Spectrometric Coronagraph \citep[LASCO;][]{brueckner1995} onboard the Solar and Heliospheric Observatory \citep[SOHO;][]{domingo1995} near Earth, but as a much fainter event to the west of the solar disc (Figure~\ref{fig:solar}(g)). The observing geometry of the CME in white light unambiguously indicates that the eruption originated from the Earth-facing Sun; however, inspection of data from the Atmospheric Imaging Assembly \citep[AIA;][]{lemen2012} instrument onboard the Solar Dynamics Observatory \citep[SDO;][]{pesnell2012} orbiting Earth reveals no clear eruptive signatures on the disc (Figure~\ref{fig:solar}(e)), even when difference images with long temporal separations are used \citep[see][]{palmerio2021b}. Nevertheless, off-limb imagery from the Extreme UltraViolet Imager (EUVI) onboard STEREO-A (Figure~\ref{fig:solar}(a)) unveils a dynamic, multi-stage eruption scenario. Specifically, the chain of events commenced with a small eruption from the northern hemisphere around 02:00~UT on June~21 (visible also on disc in AIA imagery close to N30W30), which likely caused a nearby arcade structure to lift off the Sun around 06:00~UT on the same day. This sequence of eruptions appeared in \edit1{COR1 and} COR2 imagery as an unstructured outflow, followed by an extremely faint loop-like CME that \edit1{propagated southwards and} opened the top portion of the overlying helmet streamer, diverting the remaining part towards the south. Finally, a long-lived concave-up structure reminiscent of a flux rope (indicated by an arrow in Figure~\ref{fig:solar}(a)) lifted off the southern hemisphere around 18:00~UT on June~21, \edit1{deflected towards the solar equator in the COR1 field of view,} and resulted in the streamer-blowout, three-part CME observed by COR2 (Figure~\ref{fig:solar}(c)). This structure erupted from high altitudes (its bottom was observed to lie at ${\sim}1.4\,R_{\odot}$ in EUVI imagery), possibly resulting in the lack of on-disc signatures, and was characterised by the classic ``rolling'' motion often observed in slow flux rope events. Similar events featuring extensive activity visible off the limb from one viewpoint, but without corresponding on-disc signatures, are characteristic of solar minimum conditions and were reported in several recent studies \citep{liewer2021,okane2021b}.

In order to interpret the EUVI observations and derive an approximate source region for the stealth CME, we employ a global potential field source surface \citep[PFSS;][]{wang1992} reconstruction, obtained from the low-resolution \edit3{(720$\times$360~px)} Helioseismic and Magnetic Imager \citep[HMI;][]{scherrer2012} $B_r$ synoptic map for Carrington Rotation 2232. Representative magnetic field lines of the multipolar flux system within the helmet streamer are shown in panels (b) and (f) of Figure~\ref{fig:solar} from the STEREO-A and Earth viewpoints, respectively. The complex eruptive activity observed on the STEREO-A western limb prior to and during the stealth CME eruption is evidently due to the interaction between these flux systems in a sympathetic magnetic breakout scenario \citep{torok2011,lynch2013}. The arrows in both panels point to the southern side arcade, which is the source of the larger, second eruption in response to the smaller first eruption(s) from the northern side arcade. A box encompassing the polarity inversion line (PIL) indicated in Figure~\ref{fig:solar}(f) is drawn onto the AIA image in Figure~\ref{fig:solar}(e), indicating the derived source region for the stealth CME.

Finally, in order to evaluate the geometric and kinematic parameters of the CME through the corona, we apply the Graduated Cylindrical Shell \citep[GCS;][]{thernisien2011} model to coronagraph imagery. Examples of the fitted CME are shown in panels (d) and (h) of Figure~\ref{fig:solar} over STEREO-A and SOHO data, respectively. At the time of these reconstructions, the CME apex was at ${\sim}12.5\,R_{\odot}$, its propagation direction was ($\theta$, $\phi$) = ($2^{\circ}$, $27^{\circ}$) in Stonyhurst coordinates, its axis tilt was $\gamma = -5^{\circ}$ (positive for counterclockwise rotations), and its angular width was rather modest (${\sim}30^{\circ}$ along its major axis). The CME speed obtained from successive reconstructions around this height in the corona was ${\sim}200$~km$\cdot$s$^{-1}$, i.e.\ a typical value for stealth CMEs \citep{ma2010}.

\section{OSPREI Model Setup} \label{sec:model}

The OSPREI modelling suite consists of three \edit2{coupled modules: (1)} the Forecasting a CME's Altered Trajectory \citep[ForeCAT;][]{kay2015a} model \edit2{calculates the} CME evolution through the corona especially in terms of deflections and rotations\edit2{; (2)} the Another Type of Ensemble Arrival Time Results \citep[ANTEATR;][]{kay2018} \edit2{models the heliospheric propagation and arrival time of the flux rope CME; and (3)} the ForeCAT In situ Data Observer \citep[FIDO;][]{kay2017a} produces synthetic \edit2{in-situ} magnetic field time series at the location of interest.

\begin{figure*}[!th]
\includegraphics[width=\linewidth]{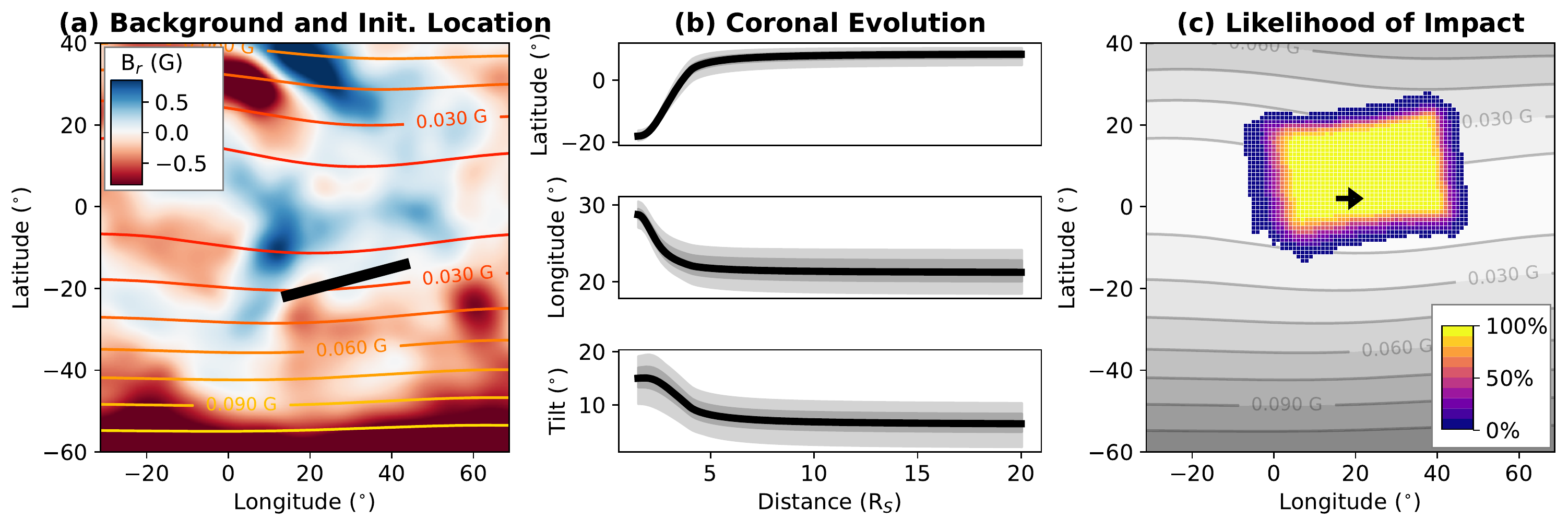}
\caption{Evolution of the 2020~June~21 stealth CME estimated with OSPREI. (a) $B_{r}$ contours (blue: positive polarity; red: negative polarity) at $1.1\,R_{\odot}$ \edit1{obtained from PFSS} using a synchronic HMI magnetogram for 2020~June~21 showing the estimated source PIL (black line). (b) ForeCAT evolution between 1.5 and $20\,R_{\odot}$. (c) ANTEATR results of the CME extent at 0.5~AU projected onto the solar surface. The projected PSP trajectory from the eruption time until the in-situ arrival of the ensemble is marked with an arrow. The contour lines in (a) and (c) indicate the absolute $B_{r}$ at the \edit1{PFSS} source surface ($2.5\,R_{\odot}$). In all panels, latitudes and longitudes are in Stonyhurst coordinates. \label{fig:forecat}}
\end{figure*}

An overview of the initial settings and resulting CME evolution modelled with OSPREI is shown in Figure~\ref{fig:forecat}. The initiation and coronal evolution \edit1{(out to $20\,R_{\odot}$)} are dealt with by ForeCAT, in which CMEs are represented as a 3D torus and their evolution is fully controlled by magnetic forces determined from a static background magnetic field---in this case, a PFSS solution on the \edit3{full-resolution (3600$\times$1440~px)} HMI synchronic magnetogram for 2020~June~21, shown in Figure~\ref{fig:forecat}(a). We initiate the CME from the PIL identified in Section~\ref{sec:solar} (marked in Figure~\ref{fig:forecat}(a)) on June~21 at 18:00~UT, with its apex at $1.5\,R_{\odot}$ based on STEREO observations of the high-altitude flux rope and with right-handed chirality based on the hemispheric helicity rule \citep{pevtsov2014}, also known as the Bothmer--Schwenn scheme \citep[from][]{bothmer1998}. To account for uncertanties in the \edit2{exact} CME source region \edit2{location}, we employ ForeCAT in its ensemble approach (with 100 members), by allowing the initial latitude and longitude to vary up to ${\pm}2^{\circ}$ and the tilt up to ${\pm}5^{\circ}$. We also allow for ensemble variations in the CME angular width ($30{\pm}5^{\circ}$ face-on and $10{\pm}1^{\circ}$ edge-on) and maximum coronal velocity ($350{\pm}25$~km$\cdot$s$^{-1}$ at $20\,R_{\odot}$). We assume a CME mass of $2{\times}10^{15}$~g and allow for variations of ${\pm}5{\times}10^{14}$~g. The CME shape is defined by the ratio of the length in the radial direction to the length in the perpendicular direction. We assume a ratio of 0.6 for the axial shape and 1.0 for the cross-section, and consider ensemble variations of 0.1 in each.

The CME's coronal evolution modelled with ForeCAT is shown in Figure~\ref{fig:forecat}(b). The black line shows the primary case, or ``ensemble seed'', the dark grey region shows the core of the ensemble distribution, and the light grey shows the full extent. \edit1{We set the CME to maintain} a constant speed of 30~km$\cdot$s$^{-1}$ until the front arrives to $4\,R_{\odot}$, at which point it begins to linearly accelerate until reaching 350~km$\cdot$s$^{-1}$ at $20\,R_{\odot}$. \edit1{This is done to emulate the slow liftoff and early evolution of the eruption in the lower corona observed by COR1 and that is characteristic of stealth CMEs \citep[e.g.,][]{palmerio2021b}.} The resulting full coronal evolution happens over ${\sim}33$~hours, which is a typical duration for streamer-blowout CMEs \citep{vourlidas2018}. Most of the CME deflections and rotations take place by ${\sim}5\,R_{\odot}$, in agreement with previous results \citep{kay2015b}. The values for latitude, longitude, and tilt ($\theta$, $\phi$, $\gamma$) evolve from ($-18^{\circ}$, $29^{\circ}$, $15^{\circ}$) to ($8^{\circ}$, $21^{\circ}$, $6^{\circ}$), with the most dramatic change experienced in latitude as the CME deflects towards the heliospheric current sheet (HCS). The ensemble shows variations of less than $5^{\circ}$ in the final position and orientation. We note that the ForeCAT values for $\theta$, $\phi$, and $\gamma$ in the outer corona are in agreement with the GCS reconstructions presented in Section~\ref{sec:solar}, \edit2{given} the expected GCS uncertainties of ${\pm}10$--$15^{\circ}$ in each parameter \citep[see][]{thernisien2009}.

Finally, Figure~\ref{fig:forecat}(c) shows the \edit2{CME's spatial} extent at 0.5~AU projected onto the \edit2{coronal source surface}. To propagate the eruption \edit1{from $20\,R_{\odot}$ outwards to PSP, we feed the ForeCAT output into ANTEATR, which \replaced{deals with}{\edit2{models}} the interplanetary propagation of CME flux ropes through a drag-inducing ambient solar wind.} Here, we use the Physics-driven Approach to Realistic Axis Deformation and Expansion version of ANTEATR \citep[ANTEATR-PARADE;][]{kay2021a}, which includes the analytical flux rope model of \citet{nieveschinchilla2018b} that is able to account for CME expansion and deformation in interplanetary space due to magnetic, thermal, and drag forces. The internal CME properties are scaled relative to the background solar values and we allow for ensemble variations in these scaling factors ($3{\pm}0.5$ for both magnetic field and temperature). ANTEATR-PARADE includes one-dimensional effects of background drag on the CME propagation speed, similar to the drag-based model of \citet{vrsnak2013}, but uses the full three-dimensional geometry to determine \edit2{the spacecraft impact and arrival time}. We set the background solar wind model using the in-situ PSP values but allow for slight variations in the velocity, number density, and magnetic field strength. Finally, our ensemble allows for variations in the CME expansion and the internal adiabatic index, as in \citet{kay2021b}.

In ANTEATR-PARADE, CMEs show a slight decrease in their face-on angular width (by ${\sim}5^{\circ}$) but maintain a roughly constant edge-on angular width. The CMEs pancake in both the axial and cross-sectional shapes, thinning in the radial direction relative to their perpendicular extent. Figure~\ref{fig:forecat}(c) shows the spatial variations in the percent chance of impact \edit2{at PSP's location} over the full ensemble using the evolved positions, orientations, shapes, and sizes. \edit2{The colour} yellow indicates \edit2{predicted} impact from all \edit2{ensemble} members. The projected trajectory of PSP from the CME eruption time until the in-situ arrival of the ANTEATR-PARADE ensemble is also indicated, suggesting a direct impact to the south of the flux rope central axis.

\section{Prediction and In-situ Comparison} \label{sec:insitu}

\begin{figure}[t!]
\includegraphics[width=\linewidth]{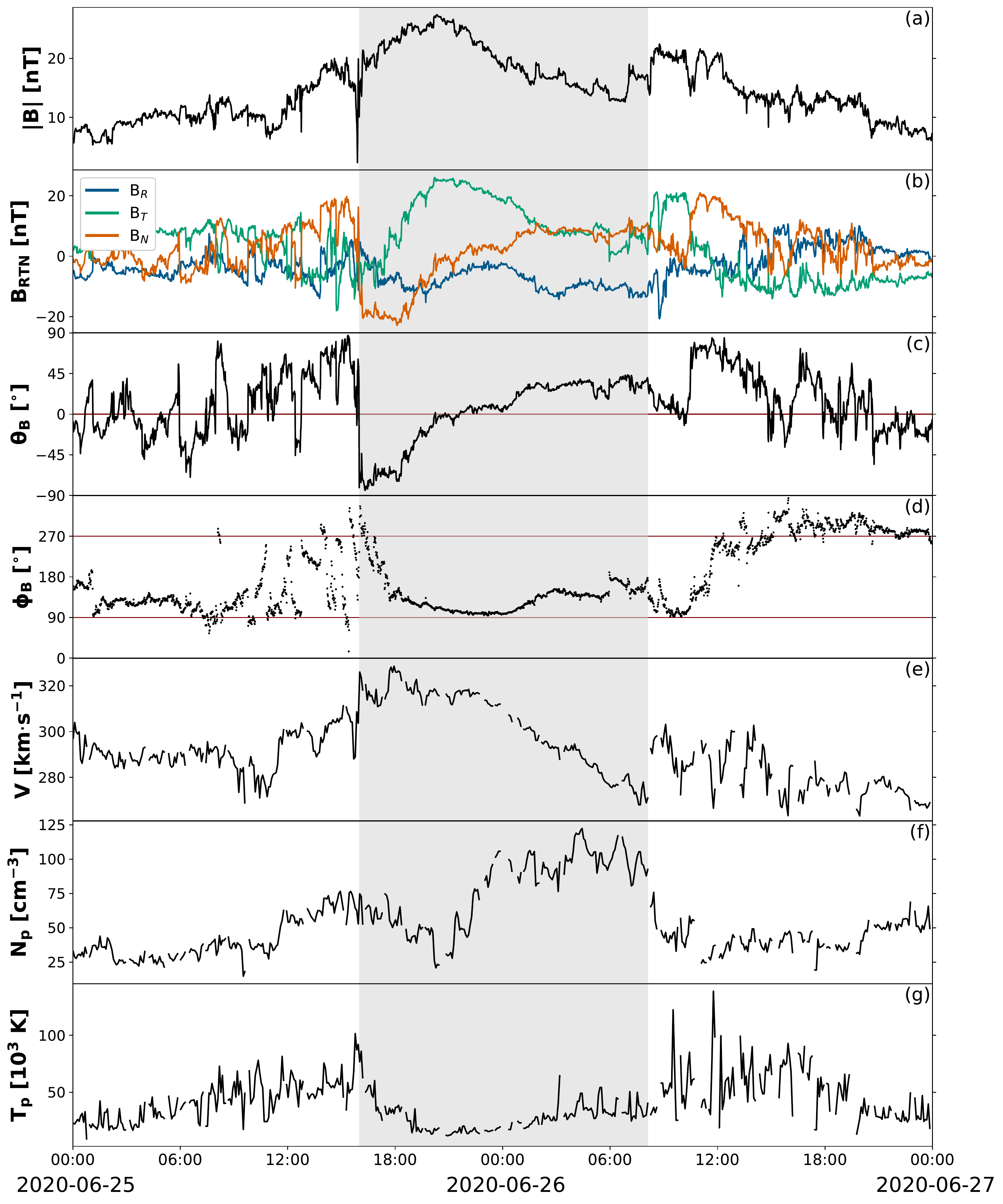}
\caption{Magnetic field and plasma measurements of the 2020~June~21 stealth CME detected in situ at PSP. The panels show, from top to bottom: (a) magnetic field magnitude, (b) magnetic field components in Radial--Tangential--Normal (RTN) Cartesian coordinates, (c) $\theta$ and (d) $\phi$ angles of the magnetic field in RTN angular coordinates, (e) solar wind speed, (f) proton number density, and (g) proton temperature. The flux rope interval is shaded in grey. \label{fig:insitu}}
\end{figure}

The 2020~June~21 stealth CME reached PSP on June~25--26, while the spacecraft was located at 0.5~AU from the Sun and ${\sim}20^{\circ}$ west of the Sun--Earth line. Figure~\ref{fig:insitu} shows in-situ measurements from the fluxgate magnetometer and Solar Probe Cup \citep[SPC;][]{case2020} instruments, part of the FIELDS \citep{bale2016} and Solar Wind Electrons Alphas and Protons \citep[SWEAP;][]{kasper2016} investigations onboard PSP, respectively. We do not find clear signatures of a CME-driven interplanetary shock, but we note a magnetic ejecta with an evident flux rope configuration. Its trailing edge is difficult to unambiguously identify based solely on magnetic field measurements, possibly because of erosion \citep[e.g.,][]{pal2020}, hence we consider here the interval associated with a descending speed profile and low proton temperature \citep[e.g.,][]{zurbuchen2006}, resulting in the shaded area in Figure~\ref{fig:insitu}. From visual inspection, the flux rope magnetic fields are characterised by a south--west--north (SWN) configuration, indicative of a low-inclination, right-handed structure.

\begin{figure*}[!ht]
\includegraphics[width=\linewidth]{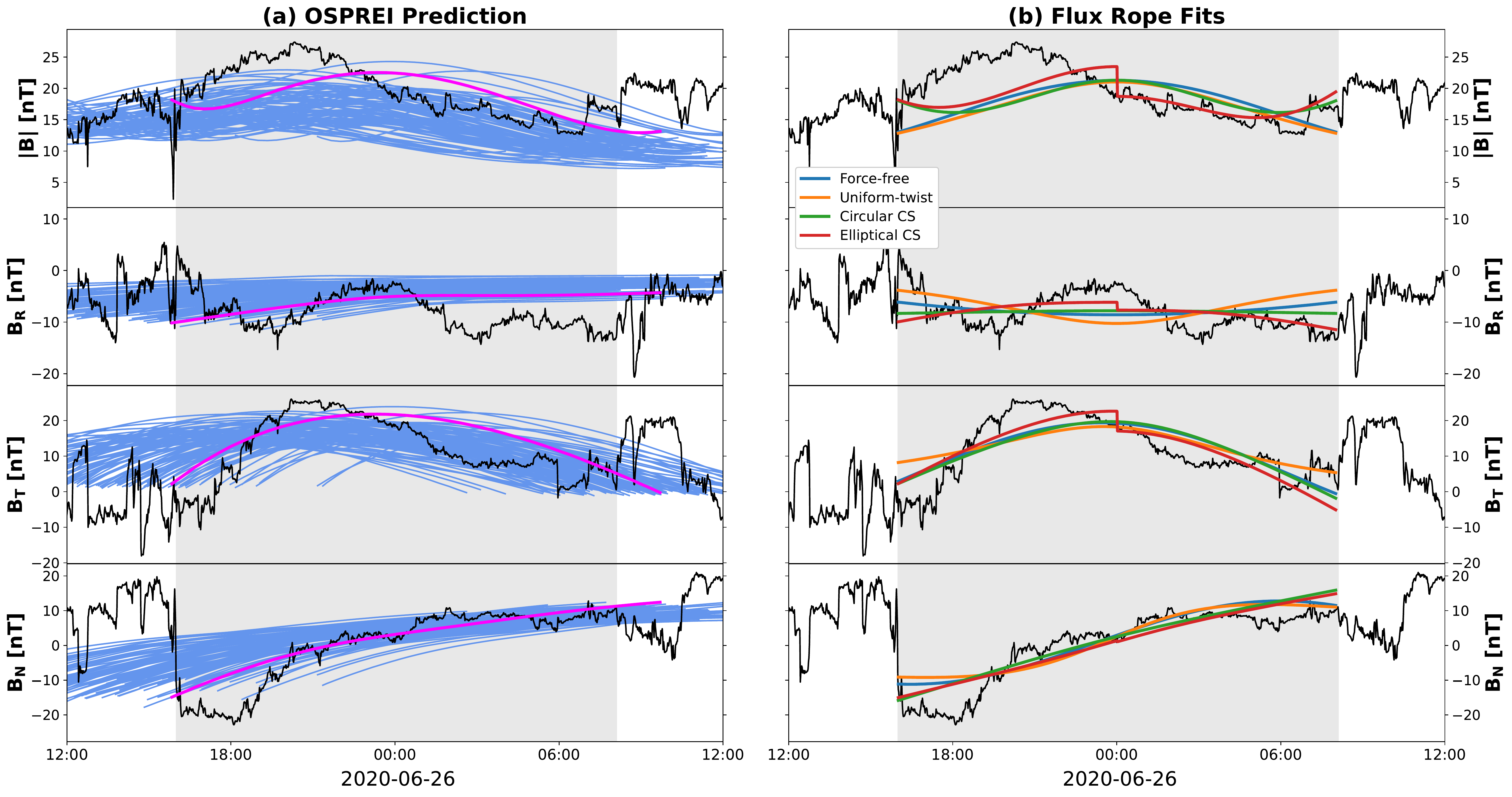}
\caption{Magnetic fields of the 2020~June~21 stealth CME compared with (a) the in-situ ensemble generated by FIDO, with the seed prediction indicated in magenta, and (b) results from four flux rope fitting techniques. The panels show, from top to bottom, magnetic field magnitude and radial, tangential, and normal components of the magnetic field in RTN coordinates.\label{fig:predvsfit}}
\end{figure*}

\edit1{To generate our prediction, we \edit2{use FIDO, i.e.} the last module of the OSPREI modelling chain, which extracts the CME position, orientation, and internal magnetic properties from ANTEATR-PARADE to create a synthetic in-situ profile corresponding to the time-dependent PSP trajectory.} \edit2{Figure~\ref{fig:predvsfit}(a) shows the OSPREI prediction ensemble in light blue and its main seed in magenta.} It is clear that the sign and large-scale sense of rotation of each magnetic field component are successfully reproduced, even when considering the full ensemble of solutions that differ in arrival time and/or amplitude depending on the evolution of the internal flux rope fields in ANTEATR-PARADE and the exact spacecraft crossing path. However, we note that the FIDO results appear less able to capture the asymmetry of the observed magnetic profiles, which are likely due to a combination of CME expansion and flux erosion, as well as local deformations of the structure. \edit2{The OSPREI ensemble seed profile predicts a flux rope orientation of ($\Theta$, $\Phi$) = ($4^{\circ}$, $75^{\circ}$) and an impact parameter of $p_0/R = 0.59$. These values correspond to a flux rope oriented almost parallel to the ecliptic plane (in agreement with the low inclination deduced from visual inspection of the PSP magnetic field data) and that is crossed below its central axis from a moderate distance, as depicted in Figure~\ref{fig:forecat}(c).} The ANTEATR-PARADE ensemble seed predicts an arrival time at PSP on 2020~June~25 at 15:50~UT, i.e.\ only 9~minutes earlier than the observation's leading edge arrival at 15:59~UT, and a flux rope passage that ends on 2020~June~26 at 09:42~UT, i.e.\ ${\sim}90$~minutes after the identified trailing edge at 08:07~UT. \edit2{The variations in arrival time \edit3{over one standard deviation of the ensemble ($\pm$3.4~hours) and the full ensemble set ($\pm$8.4~hours)} are well within the current CME (or CME-driven shock) arrival time uncertainties \citep[of the order of $\pm$10~hours; e.g.,][]{riley2018}.}

In order to \edit2{further} interpret and evaluate the \edit2{FIDO} in-situ magnetic field profiles with respect to PSP measurements, we \edit2{also} compare \edit2{the predicted ensemble seed flux rope orientation} with \edit2{results from} a number of \edit2{commonly used in-situ} flux rope \edit2{models applied to PSP data}. \edit2{We implement the} constant-$\alpha$ force-free \edit2{cylinder model} \citep{lepping1990}, \edit2{the} uniform-twist \edit2{model} \citep{farrugia1999}, \edit2{and the non-force free} circular cross-section \citep{hidalgo2000} and elliptical cross-section \citep{hidalgo2002b} flux rope \edit2{models}. Since it has been shown that different flux rope reconstructions can yield significantly different results \citep{riley2004,alhaddad2013}, the use of multiple \edit2{fitting} techniques ensures a more robust estimation of the \edit2{large-scale, coherent} magnetic \edit2{structure} of the 2021~June~21 CME at PSP. \edit2{The flux rope parameters for each of the in-situ model reconstructions are compared with the OSPREI prediction in Table~\ref{tab:fits} and the magnetic field profiles obtained from each fit are shown in Figure~\ref{fig:predvsfit}(b).} The flux rope fits shown in Table~\ref{tab:fits} all yield a right-handed structure with a low-inclination central axis that is directed westward, in agreement with the SWN configuration retrieved from visual inspection \edit2{and the FIDO results. The values shown in Table~\ref{tab:fits} yield a mean inclination angle of $\Theta = 6.0{\pm}2.5^{\circ}$, a mean azimuthal angle of $\Phi = 87.6{\pm}19.4^{\circ}$, and a mean impact parameter of $p_0/R = 0.43{\pm}0.24$. The standard deviations in each of these parameters range from significantly less than to approximately the same as the statistical uncertainties derived in the constant-$\alpha$ force-free fitting procedure \citep[e.g.,][]{lepping2003,lynch2005a}. Since the in-situ flux rope modelling techniques show considerable agreement, the CME flux rope under analysis can be deemed a ``simple'' case \citep[see][]{alhaddad2018}, for which fitting results should be considered more reliable.} 

\begin{table}[t!]
\centering
\caption{OSPREI prediction and fitting results for different techniques applied to PSP in-situ measurements of the 2020~June~21 stealth CME. The parameters shown are: orientation angles of the flux rope symmetry axis ($\Theta$, $\Phi$) in RTN coordinates, spacecraft impact parameter ($p_{0}$), normalised goodness-of-fit measure ($\chi^{2}$), \edit1{and quality score ($\epsilon$)}.}
\label{tab:fits}
\hspace*{-0.24\columnwidth}
\begin{tabular}{|l|ccccc|}
\tableline\tableline
 & $\Theta$ [$^{\circ}$] & $\Phi$ [$^{\circ}$] & $p_{0}$ [$R^{-1}$] & $\chi^{2}$ & $\epsilon$ \\[0.02in]  
\tableline
OSPREI  & $4$ & $75$ & $0.59$ & $0.27$ & $0.38$ \\
Force-free  & $9$ & $79$  & $0.48$ & $0.18$ & $0.35$ \\
Uniform-twist  & $7$ & $121$ & $0.02$ & $0.23$ & $0.39$ \\
Circular CS  & $7$ & $88$  & $0.45$ & $0.17$ & $0.34$ \\
Elliptical CS & $3$ & $75$ & $0.60$ & $0.15$ & $0.31$ \\[0.02in]
\tableline\tableline
\end{tabular}
\end{table}

\edit1{Finally, in Table~\ref{tab:fits} we also compare two error metrics for each model. The goodness-of-fit measure $\chi^{2}$ is the standard metric used to quantify discrepancies between a model and data in the rotation of the magnetic field unit vector \citep[e.g., ${\chi}^{2}_{\mathrm{dir}}$ in][]{lynch2003}, whilst the quality score $\epsilon$ quantifies differences in the field magnitudes \citep[e.g., $\delta$ in][]{kay2017c}. The similarity of the values in each error metric between the flux rope fits and the OSPREI prediction demonstrate that, at least in this ``simpler'' case of a slow, streamer-blowout CME, magnetic field estimates \edit2{that are forward-modelled} using uniquely solar observations as input perform comparably to \edit2{flux rope} fitting \edit2{of in-situ data}.}

\section{Discussion and Conclusions} \label{sec:conclusions}

In this \edit3{work}, we have estimated the magnetic fields of a stealth CME that erupted on 2020~June~21 and was observed in situ by PSP at 0.5~AU on 2020~June~25--26. After locating the CME source region with the aid of off-limb STEREO observations and PFSS reconstructions, we have \edit2{forward-}modelled the event from its eruption to its detection at PSP using the OSPREI modelling suite. We have then compared the resulting magnetic field profiles with in-situ measurements of the CME as well as flux rope fitting results, finding \edit2{agreement (and arguably considerable agreement)} for all components. From a forecasting perspective, successfully capturing $B_{T}$ and $B_{N}$ is the most critical aspect in determining the magnetic configuration of a flux rope and, hence, its space weather effects \citep[e.g.,][]{kilpua2019a,temmer2021}. Nevertheless, the OSPREI prediction possibly accounted for less expansion than in the real case, resulting in a more symmetric structure than was observed by PSP. Asymmetries and distortions of the flux rope body, whilst not necessarily paramount for forecasting purposes, are important to better understand the internal structure and evolution of CMEs, hence future improvements in modelling capabilities will require greater characterisation of these processes. As PSP approaches the Sun with each perihelion pass, the opportunity to encounter CMEs at smaller and smaller heliocentric distances may provide valuable data and physical constraints to advance in this direction.

\edit1{It is worth commenting on two assumptions that have been employed in this work. Firstly, we have used the ``classic'' PFSS source surface height of $2.5\,R_{\odot}$, which has however been challenged during the PSP era \citep{bale2019,riley2019} and more generally during solar minimum \citep{lee2011}. Further development of the ForeCAT ensemble modelling could incorporate a range of source surface heights.} Secondly, we have assumed the erupting flux rope chirality based on the hemispheric helicity rule, which however holds true only in ${\sim}75$\% of cases \citep{pevtsov2014}. Future efforts in predicting the magnetic fields of stealth CMEs may avail themselves of nonlinear force-free magnetic field extrapolations from photospheric vector field observations in order to obtain an ``independent'' measure of the helicity starting from an approximate source region on the Sun \citep{wiegelmann2008}. However, these techniques are best suited for strong-field active region configurations and may not be especially useful for quiet-Sun or decayed active region sources. Given that stealth CMEs tend to originate at higher altitudes and over larger scales, global evolutionary modelling via surface flux transport or magnetofrictional relaxation may be better suited for determining the helicity of extended quiet-Sun PILs \citep[e.g.,][]{mackay2014,yardley2021}.

Finally, the event studied here is an excellent example of the classic slow, streamer blowouts that constitute the main CME population during solar minimum. The encouraging results of the magnetic field \edit2{hindcast} prediction presented in this work suggest that these events could be accurately forecast even if their exact source region remains elusive in on-disc imagery. Certainly, in this and other reported cases off-limb observations were crucial in connecting a stealth CME at the Sun with its interplanetary counterpart \citep[e.g.,][]{korreck2020,lario2020,okane2021b}. Thus, this study highlights the importance of continuous observations of the solar disc and corona away from the Sun--Earth line, not only to enable more accurate space weather forecasts, but also to shed light on the complex characteristics and dynamics of stealthy CMEs that, albeit slow, have potential for considerable geoeffectiveness.


\begin{acknowledgments}
E.P. is supported by the the NASA Living With a Star Jack Eddy Postdoctoral Fellowship Program, administered by UCAR’s Cooperative Programs for the Advancement of Earth System Science (CPAESS) under award no. NNX16AK22G.
C.K. is supported by NASA under grants no. 80NSSC19K0274, issued through the Heliophysics Guest Investigators Program, and no. 80NSSC19K0909, issued through the Heliophysics Early Career Investigators Program.
N.A. and W.Y. acknowledge the support of NSF AGS-1954983 and NASA 80NSSC21K0463.
B.J.L. acknowledges support from NSF AGS-1622495, NASA NNX17AI28G, and NASA 80NSSC19K0088.
S.P. acknowledges the European Research Council (ERC) under the European Union's Horizon 2020 Research and Innovation Program Project SolMAG 724391.
C.O.L. acknowledges support from AFOSR Grant FA9550-16-1-0418.
\end{acknowledgments}

\vspace{5mm}
\facilities{PSP (FIELDS, SWEAP); SDO (AIA, HMI); SOHO (LASCO); STEREO (SECCHI)}
\software{OSPREI; SolarSoft \citep{freeland1998}; SunPy \citep{sunpy2020}}

\newpage

\bibliography{bibliography}{}
\bibliographystyle{aasjournal}

\end{document}